\documentclass[aps,prd,twocolumn,superscriptaddress,nofootinbib,showpacs]{revtex4}

\usepackage{graphicx}
\usepackage{amsmath,amssymb,latexsym}
\usepackage{bm}
\usepackage{epsfig}

\newcommand{\postscript}[2]{\setlength{\epsfxsize}{#2\hsize}
   \centerline{\epsfbox{#1}}}

\begin{document}



\title{AMANDA Observations Constrain the Ultra-High Energy Neutrino Flux}

\author{Francis Halzen}
\affiliation{Department of Physics,
University of Wisconsin, Madison, WI}

\author{Dan Hooper} \affiliation{Particle Astrophysics Center,
  Fermilab, Batavia, IL}

\begin{abstract}
  \noindent A number of experimental techniques are currently being deployed in an effort to make the first detection of ultra-high energy cosmic neutrinos. To accomplish this goal, techniques using radio and acoustic detectors are being developed, which are optimally designed for studying neutrinos with energies in the PeV-EeV range and above. Data from the AMANDA experiment, in contrast, has been used to place limits on the cosmic neutrino flux at less extreme energies (up to $\sim$10 PeV). In this letter, we show that by adopting a different analysis strategy, optimized for much higher energy neutrinos, the same AMANDA data can be used to place a limit competitive with radio techniques at EeV energies. We also discuss the sensitivity of the IceCube experiment, in various stages of deployment, to ultra-high energy neutrinos.

\end{abstract}

\pacs{95.55.Vj, 95.85.Ry, 98.70.Sa; MADPH-06-1465; FERMILAB-PUB-06-097-A}

\maketitle

As ultra-high energy protons and nuclei propagate over cosmological distances, they interact with the cosmic microwave and infra-red backgrounds, producing pions~\cite{gzk}. These pions decay, generating neutrinos with typical energies in the range of $10^7$ to $10^{10}$ GeV. These ultra-high energy particles, known as cosmogenic or GZK neutrinos, have not yet been observed.

A wide range of experimental efforts are currently underway to detect ultra-high energy neutrinos \cite{review}. These include experiments using radio antennas such as RICE under the Antartic ice of the South Pole~\cite{rice}, and the balloon mission ANITA (and its predecessor ANITA-lite)~\cite{anita}. Techniques for observing ultra-high energy neutrinos using acoustic detectors are also being explored, although the prospects for this technology are not yet well understood \cite{acoustic}. Cosmic ray experiments, such as the Pierre Auger Observatory~\cite{auger}, are also capable of detecting ultra-high energy neutrinos by observing slighly upgoing showers produced by Earth-skimming tau neutrinos or deeply penetrating quasi-horizontal showers~\cite{earthskimming}.

Each of these experimental techniques is specifically suited to studying particles ({\it ie.}~neutrinos or cosmic rays) at extremely high energies. The effective volumes of RICE, AUGER and ANITA become appreciable only above roughly $\sim$$10^7$~GeV, $\sim$$10^8$~GeV and $\sim$$10^9$~GeV, respectively. Acoustic techniques are likely to have energy thresholds comparable to or higher than these other methods. In contrast, high-energy neutrino telescopes using optical detectors, such as AMANDA \cite{cascadelimit}, ANTARES \cite{antares}, IceCube \cite{icecube}, and KM3 \cite{km3}, while also capable of observing ultra-high energy neutrinos \cite{jaime}, are designed to be sensitive to neutrino induced cascades with energies as low as a few TeV.

The strongest limit on contained neutrino-induced cascades in the $\sim$PeV energy region has been published by the AMANDA collaboration \cite{cascadelimit}. This limit ($E^2_{\nu} dN_{\nu}/dE_{\nu} < 8.6 \times 10^{-7}$ GeV s$^{-1}$ sr$^{-1}$ cm$^{-2}$, at the 90\% confidence level, in the 50 TeV to 6 PeV energy range) was arrived at after making a number of experimental cuts on the data to remove a large fraction of the background from atmospheric neutrinos and muons. Furthermore, the analysis was optimized for a spectrum falling with $dN_{\nu}/dE_{\nu} \propto E_{\nu}^{-2}$.

This approach is less than ideal for searching for cosmogenic neutrinos, which appear as isolated spectacular events of much greater energy ($\sim$10 EeV). At these energies, the showers generated in an experiment such as AMANDA are very large (approximate radii of $\sim 300$ meters for a $10^7$ GeV shower and $\sim 500$ meters for a $10^{11}$ GeV shower; considerably larger than the instrumented width of AMANDA). Such enormous volume events are expected to have a negligibly small background due to the rapidly falling spectrum of atmospheric neutrinos and muons. The information contained in these events is so much superior to that of average AMANDA events, that throughout this study we assume that showers and muons can be efficiently seperated. Given that the background can be effectively eliminated, the optimal analysis strategy for studying this class of events is very different from that suited for neutrinos with TeV-PeV energies.

The largest shower reconstructed in the AMANDA experiment has an energy close to 100 TeV, occupying a volume that triggered fewer than 300 of its 677 modules \cite{largest}. Motivated by the fact that no larger events have been observed, we have considered an alternative cut designed to retain as many potential ultra-high energy events as possible. This cut is to simply consider only events in which 300 or more of the modules trigger while omitting the restrictive cuts of Ref.~\cite{cascadelimit}. Our simulation adopts a simple geometric approach in which spherical showers are overlayed against the AMANDA geometry (roughly a cylinder with height and diameter of 600 and 200 meters, respectively) \cite{icecubeplus}. This method reflects the science involved and can be considered a good approximation for such extremely large events, even in the absence of a full and detailed detector simulation. The physics of neutrino interactions and detection at these high energies has been established and confirmed though the calibration of the AMANDA experiment.

\begin{figure}
 \postscript{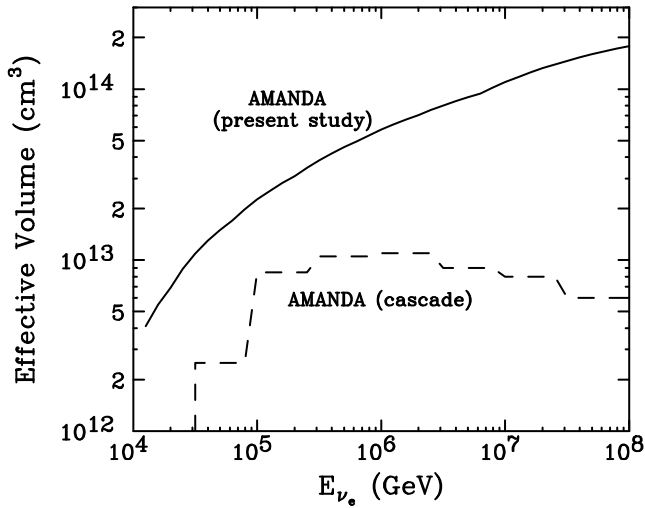}{0.98} 
  \caption{The effective volume of AMANDA to electron neutrinos as found using the cuts from the AMANDA cascade analysis \cite{cascadelimit} and using the $>$300 module cut adopted in this study.}
\label{vol}
\end{figure}

\begin{figure}
 \postscript{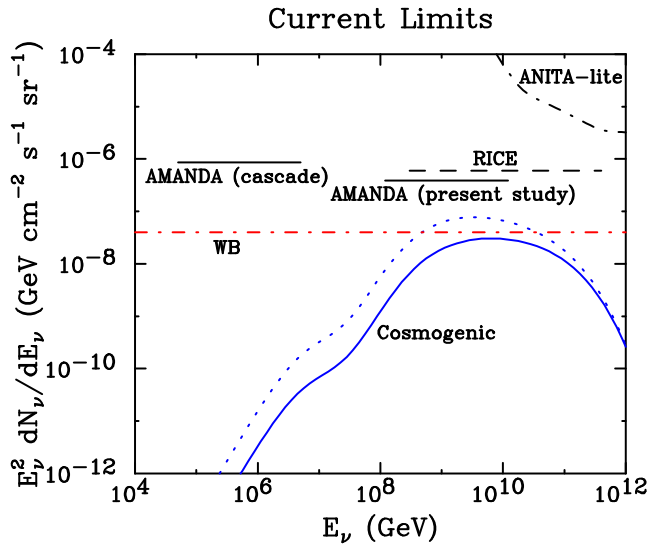}{0.98} 
  \caption{The current experimental limits on the diffuse flux of high and ultra-high energy neutrinos (summed over flavors in the ratio $\phi_{\nu_e}:\phi_{\nu_{\mu}}:\phi_{\nu_{\tau}}=1:1:1$, after oscillations are considered). The AMANDA cascade, RICE and ANITA-lite limits are found in Refs.~\cite{cascadelimit}, \cite{rice} and \cite{anitalite}, respectively. The AMANDA and ANITA-lite limits are shown at the 90\% confidence level, while the RICE limit shown is at 95\% confidence. For comparison, we also show the neutrino spectrum predicted from the propagation of ultra-high energy cosmic ray protons (the cosmogenic flux)~\cite{cosmogenic} and the Waxman-Bahcall bound~\cite{wb}. The solid and dotted curves correspond to cosmogenic neutrino fluxes calculated assuming normal ($\propto (z+1)^3$) and strong ($\propto (z+1)^4$) source evolution, respectively.}
\label{limit}
\end{figure}

\begin{figure}
 \postscript{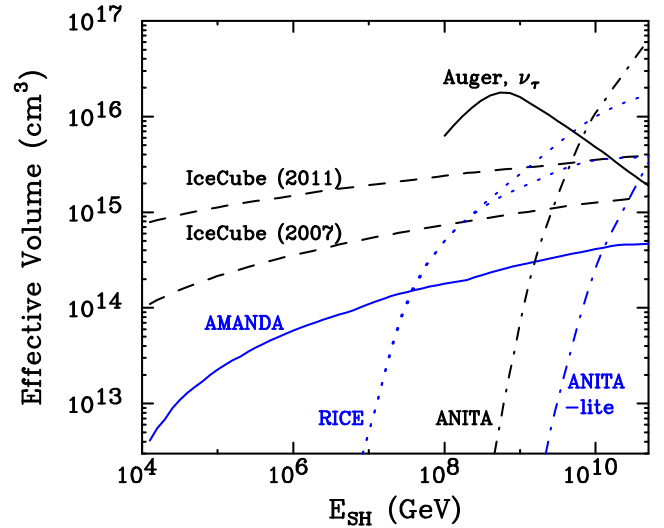}{0.98} 
  \caption{The effective volumes (water equivalent) of various current and near future ultra-high energy neutrino detectors. The volume of the AMANDA experiment has been calculated using the $>$300 module cut described in the text. A similar approach was taken for IceCube, requiring a number of modules be triggered equivalent to a volume of 300 modules in AMANDA (the module spacing in IceCube is greater than in AMANDA). The upper and lower RICE curves (dots) correspond to hadronic and electromagnetic showers, respectively. The ANITA, ANITA-lite and Auger curves are shown for comparison with AMANDA, IceCube and RICE in terms of $2\pi$ steridians equivalent. In the case of ANITA and ANITA-lite, the volumes have been scaled by a factor of 10/365 to account for the 10 day duration of each balloon mission. See text for more details.}
\label{volfut}
\end{figure}

In Fig.~\ref{vol}, we compare the effective volume (for electron neutrinos) of the AMANDA experiment, as calculated using the cuts adopted in the cascade analysis optimized for $\sim$PeV energies~\cite{cascadelimit}, to the effective volume using our $>$300 module cut. Our cut would remove far fewer of the signal events. Whereas it may be debatable how well background rejection at energies below $\sim$PeV would work, we expect the sample to have very low background from atmospheric neutrinos at GZK energies.

So far, 6 years of AMANDA data has been analyzed (with approximately 200 live days per year). Assuming that no events with more than 300 triggered modules will be found in this sample, it can be used to place a limit on the flux of ultra-high energy cosmic neutrinos. In Fig.~\ref{limit}, we plot this limit and compare it to the current limit of the RICE \cite{rice} and ANITA-lite \cite{anitalite} experiments and the AMANDA cascade analysis limit~\cite{cascadelimit}. From this figure, it is clear that AMANDA, even though capable of studying neutrinos of lower energies, is highly competitive as an ultra-high energy neutrino detector. In the $10^8$-$10^{10}$ GeV range, the AMANDA limit we find here is comparable to the limit placed by RICE~\cite{rice}. Also shown in Fig.~\ref{limit} are the predictions for the cosmogenic neutrino flux generated in the propagation of ultra-high energy protons for the cases of normal (solid) and strong (dots) source evolution~\cite{cosmogenic}. The Waxman-Bahcall bound is also shown for comparison (dot-dash) \cite{wb}. By noticing that these benchmark neutrino flux models are not far below the current limits of AMANDA and RICE, we assess that as these experiments continue to collect data, they may detect the first ultra-high energy neutrino event at any time.

Over the next few years, a number of experiments will become considerably more sensitive to neutrinos with ultra-high energies. The first ANITA flight is scheduled to take place later this year, and the Pierre Auger Observatory continues to be constructed, hopefully to be completed by 2007. The first 9 (out of 80) strings of IceCube modules are currently operating, with another 12-14 scheduled to be deployed in the 2006-2007 season. Its full cubic kilometer instrumented volume should be completed around 2011. Future deployment of IceCube strings will also be accompanied by additional radio anntennas, further extending the volume of RICE.

\begin{table*}
\begin{tabular}{|c@{}|c|c@{}|c|c|}
  \hline
  \cline{1-3} \cline{4-5}
  & ~~Event Rate~~ & Current Exposure & 2008 Exposure  & 2011 Exposure~~ \\
  \hline
  \hline
  AMANDA (300 hits) & 0.044 yr$^{-1}$ & 3.3 yrs, 0.17 events & NA & NA \\
  \hline
  IceCube, 2007 (300 hits equiv.) & 0.16 yr$^{-1} $& NA & 0.4 events  & NA\\
  \hline
  IceCube, 2011 (300 hits equiv.) & 0.49 yr$^{-1} $& NA & NA & 1.2 events\\
  \hline
  \hline
  RICE & $\sim 0.07$ yr$^{-1} $& 2.3 yrs, 0.1-0.2 events  & 0.2-0.3 events  & 0.3-0.4 events\\
  \hline
  \hline
  ANITA-lite & 0.009 per flight \cite{anitalite} & 1 flight, 0.009 events & NA & NA\\
  \hline
  ANITA & $\sim 1$ per flight & NA & 1 flight, $\sim 1$ event & 3 flights, $\sim 3$ events \\
  \hline
  \hline
  Pierre Auger Observatory & 1.3 yr$^{-1}$ \cite{augervol} & NA & $\sim 2$ events & $\sim 5$ events \\
  \hline
  \hline
\end{tabular}
\caption{Estimated event rates in various ultra-high energy neutrino experiments assuming the standard cosmogenic neutrino flux of Ref.~\cite{cosmogenic}. Projected exposures have assumed 200 live days per year for AMANDA/IceCube and RICE. The rates given for AMANDA and IceCube account for cascade events only, and do not include muons from $\nu_{\mu}$ charged current interactions. Over the next few years, each of these four experimental programs will reach the exposure needed to potentially observe the first ultra-high energy neutrino.}
\label{rates}
\end{table*}

In Fig.~\ref{volfut}, we compare the effective volumes ($2\pi$ steredians, water equivalent) of several current and future ultra-high energy neutrino experiments. The effective volumes shown for AMANDA and IceCube have been calculated using the same $>$300 module cut as described earlier in this letter (in the case of IceCube, we have used a volume cut equivalent to that of 300 modules in AMANDA. The actual number of modules required is modified by the string spacing being greater in IceCube than in AMANDA). By IceCube (2007), we refer to a configuration of 23 IceCube strings, as they may be arranged following the 2006-2007 deployment season. By IceCube (2011), we refer to the completed IceCube configuration with all 80 strings.

The effective volume of RICE shown in Fig.~\ref{volfut} (shown as a dotted line) seperates into two lines above $\sim 10^8$ GeV. The higher and lower curves correspond to the effective volumes for hadronic and electromagnetic showers, respectively~\cite{rice}. For the effective volumes shown for the ANITA \cite{anitavol}, ANITA-lite and Pierre Auger experiments, we have implicitly included the solid angle and time windows of these experiments for more convienent comparison. The volumes shown are equivalent to a $2\pi$ solid angle field of view and, for ANITA and ANITA-lite, a factor of $10/365$ was included to compare the exposure of a ten day flight to a year of operation by AMANDA, IceCube, RICE or Auger. For the Pierre Auger Observatory, we show the approximate equivalent effective volume for Earth-skimming tau neutrino induced cascades. The exact shape of this contour is complicated by the effects of tau absorption and regeneration in the Earth. For more details, see Ref.~\cite{augervol}.

In table~\ref{rates}, we estimate the event rates in each of these experiments, for the case of the cosmogenic neutrino flux ($n=3$)~\cite{cosmogenic}. With the current integrated exposure, AMANDA and RICE each expect 0.1-0.2 events from such a flux. By the end of 2008, they will collectively expect about $\sim$0.7 events. By 2011, IceCube alone will have accumulated the exposure needed to observe $\sim$1.2 events, although the precise value will depend on the configuration of strings in various intermediate stages of deployment.\footnote{The rates for AMANDA and IceCube shown in the table are for cascade events alone and do not include the rate of muons produced in charged current $\nu_{\mu}$ interactions. The rate of muons is expected to be similar to that of cascades~\cite{icecubeplus}.} Each ANITA flight is expected to observe of the order of 1 cosmogenic neutrino event, although this rate depends critically on the precise energy threshold of the experiment. The Pierre Auger Observatory, currently under construction at its Southern cite in Argentina, is expected to be completed by 2007. They also anticipate of the order of 1 cosmogenic neutrino event per year, generated by Earth-skimming tau neutrinos\cite{augervol}.

The standard cosmogenic neutrino flux we have used as our benchmark scenario may plausibly underestimate the rates observed in these experiments. If the source evolution scales with $(z+1)^4$ rather than a homogeneous distribution scaling as $(z+1)^3$, the flux of neutrinos will be larger, especially below $\sim 10^{10}$ GeV \cite{cosmogenic,strong}. With such a flux, rates for AMANDA and IceCube will be larger by a factor of $\sim$3 than those given in table~\ref{rates}. RICE and Auger will benefit somewhat less from strong source evolution, as they are most sensitive at higher energies where the effects of strong source evolution are less pronounced. ANITA's rates will be increased only slightly as a result of strong source evolution.

On the other hand, the cosmogenic neutrino flux could be smaller if the observed ultra-high energy cosmic rays are generated in local sources, or if they consist of heavy or intermediate mass nuclei rather than protons~\cite{heavy}.

Experiments beyond those discussed thus far in this letter will likely be capable of detecting far larger numbers of ultra-high energy neutrinos. A  northern Pierre Auger Observatory site, currently in the planning phase, could potentially be expanded to a surface area several times larger than its southern counterpart. Extensions of IceCube, optimized for ultra-high energies, have also been proposed \cite{icecubeplus}. Very large acoustic and radio arrays are also promising~\cite{acoustic}. The possibility of using natural rock salt, rather than water or ice, as a medium for ultra-high energy neutrino detection is also being explored \cite{rocksalt}.

In summary, although it is also capable of observing neutrino induced cascades with energies as low as a few TeV, we demonstrate that AMANDA is also very competative as an ultra-high energy experiment. Assuming that no very large volume cascades are present within the six years of analyzed AMANDA data, we calculate a limit on the ultra-high energy neutrino flux. This limit is comparable to that published by the RICE collaboration, and is considerably stronger than the limit from ANITA-lite. 

As the IceCube experiment continues to be deployed, the exposure of the AMANDA/IceCube program is accumulating at an increasing rate. At the end of 2008, $\sim$0.4 events will be expected at AMANDA/IceCube, for the case of a standard cosmogenic neutrino flux. By the time of the completion of IceCube in 2011, the accumulated exposure is expected to generate $\sim$1.2 events. This is competative with the rates anticipated from RICE, ANITA and the Pierre Auger Observatory. Given this wide range of technologies, each nearing the sensitivity needed to study the predicted cosmogenic neutrino flux, the first detection of an ultra-high energy neutrino may occur at any time.

\smallskip

We would like to thank Dave Besson and Amy Connolly for helpful discussions. DH is supported by the US Department of Energy and by NASA Grant NAG5-10842.

\end{document}